\documentclass[10pt,journal,compsoc]{IEEEtran}

\ifCLASSOPTIONcompsoc                  % if we use pdflatex
  \pdfoutput=1\relax                   % create PDFs from pdfLaTeX
  \usepackage{graphicx}                % allow us to embed graphics files
  \DeclareGraphicsExtensions{.pdf,.png,.jpg,.jpeg} % for pdflatex we expect .pdf, .png, or .jpg files
\else%                                 % else we use pure latex
  \usepackage{graphicx}                % allow us to embed graphics files
  \DeclareGraphicsExtensions{.eps}     % for pure latex we expect eps files
\fi%

\graphicspath{{figures/}{pictures/}{images/}{./}} % where to search for the images
\usepackage{float}
\usepackage{microtype}                 % use micro-typography (slightly more compact, better to read)
\PassOptionsToPackage{warn}{textcomp}  % to address font issues with \textrightarrow
\usepackage{textcomp}                  % use better special symbols
\usepackage{mathptmx}                  % use matching math font
\usepackage{times}                     % we use Times as the main font
\usepackage{cite}                      % needed to automatically sort the references
\usepackage{tabu}                      % only used for the table example
\usepackage{booktabs}                  % only used for the table example
\usepackage{enumitem}

\begin{document}
\title{ClinicLens: Visual Analytics for Exploring and Optimizing the Testing Capacity of Clinics given Uncertainty}

\author{Yu~Dong,
        Jie~Liang,
        Longbing~Cao,~\IEEEmembership{Senior~Member,~IEEE,}
        and Daniel~Catchpoole% <-this % stops a space
\IEEEcompsocitemizethanks{\IEEEcompsocthanksitem Yu Dong, Christy Jie Liang, and Longbing Cao are with the School of Computer Science, University of Technology, Sydney, Australia.\protect\\

%\IEEEcompsocthanksitem LongBing Cao was with the Data Science Institute, University of Technology, Sydney, Australia. 
%\protect \\
%E-mail: Longbing.Cao@uts.edu.au. \protect \\
\IEEEcompsocthanksitem Daniel Catchpoole is with the Faculty of Engineering and Information Technology, University of Technology, Sydney, and the Children’s Hospital at Westmead, Australia.\protect\\% <-this % stops a space
\IEEEcompsocthanksitem E-mail: \{Yu.Dong,Jie.Liang,Longbing.Cao,Daniel.Catchpoole\}@uts.edu.au.} \protect \\
\thanks{Manuscript received April 19, 2005; revised August 26, 2015.}}


\markboth{IEEE TRANSACTIONS ON VISUALIZATION AND COMPUTER GRAPHICS, October~2022}, %
\IEEEtitleabstractindextext{%
\begin{abstract}
Clinic testing plays a critical role in containing infectious diseases such as COVID-19. However, one of the key research questions in fighting such pandemics is how to optimize testing capacities across clinics. In particular, domain experts expect to know exactly how to adjust the features that may affect testing capacities, given that dynamics and uncertainty make this a highly challenging problem. Hence, as a tool to support both policymakers and clinicians, we collaborated with domain experts to build \textit{ClinicLens}, an interactive visual analytics system for exploring and optimizing the testing capacities of clinics. \textit{ClinicLens} houses a range of features based on an aggregated set of COVID-19 data. It comprises Back-end Engine and Front-end Visualization that take users through an iterative exploration chain of extracting, training, and predicting testing-sensitive features and visual representations. It also combines AI4VIS and visual analytics to demonstrate how a clinic might optimize its testing capacity given the impacts of a range of features. Three qualitative case studies along with feedback from subject-matter experts validate that \textit{ClinicLens} is both a useful and effective tool for exploring the trends in COVID-19 and optimizing clinic testing capacities across regions. The entire approach has been open-sourced online: \textit{https://github.com/YuDong5018/clinic-lens}.

\end{abstract}

}

\maketitle

% is not selected <OR> if conference mode is selected - because compsoc
% conference papers position the abstract like regular (non-compsoc)
\IEEEdisplaynontitleabstractindextext

% \fi
\IEEEpeerreviewmaketitle

\ifCLASSOPTIONcompsoc
\IEEEraisesectionheading{\section{Introduction}\label{sec:introduction}}
\else

\section{Introduction}
\label{sec:introduction}
\fi

\IEEEPARstart{T}{he} worldwide COVID-19 pandemic has caused severe strain on healthcare systems and has had a detrimental impact on society and economies all over the world \cite{cao2022covid}. To date, growing attention has been paid to visualization techniques, such as visual dashboards \cite{samet2020using,usman2022multi,jiang2020interactive}, CT diagnoses \cite{jadhav2021covid,liu2021medical}, and genomics modeling \cite{lv2021detection,lyi2021gosling} as a way of helping domain experts comprehend, analyze, and model COVID-19 \cite{cao2022covid}. As such, visual analytics is helping to contain and control this global crisis \cite{zhang2021mapping}. 

Information visualization is playing a crucial role in exploring COVID-19 at macro-levels. Already, a vast number of visual dashboards have been implemented to support correlation analysis tasks both post-hoc and in real-time \cite{zhang2022visualization}. These dashboards typically aggregate COVID-19-related information into visualizations that can be viewed from multiple perspectives. However, the data analyzed tend to be limited. For example, a tool might show the current status of cases across countries \cite{WHO2022dashboard,dong2020interactive} or the broad case trends within a country \cite{Au2022statistics,NSW2022statistics}. Additionally, the sheer scale of the COVID-19 data available for analysis is showing us just how insufficient our information visualization tools really are. COVID-19 data contains many different data types, including in-depth research, and the data gathered are generally characterized by a high amount of uncertainty. This typically makes analyzing COVID-19 trends a very complex task. Hence, if we are to better manage COVID-19 and future infectious disease outbreaks, more research into the nature of the information itself is required \cite{porter2022new,cao2022covid}. For example, we might look to incorporate more metadata into our datasets where spatiotemporal features allocate beside multidimensional features.  This is a need that has been listed in a systematic survey as a valuable future direction of research to prevent the spread of disease \cite{deng2023survey}.  In these situations, despite the introduction of adequate visual analytics methods \cite{leite2020covis,yang2022epimob,yu2022user} and AI4VIS \cite{naseem2021covidsenti,padilla2022multiple,rezaee2022smart} with the goal of providing interactions and artificial intelligence advantages to deal with data correlations and other complexities, dynamic and multi-aspect visual analyses of COVID-19 remain insufficient.

Recapping the last three years, one valuable lesson learned from COVID-19 is the vital role of clinic engagement in responding to the pandemic \cite{national2020covid}. During this period, numerous clinics were involved in sampling, testing, and diagnosing the disease in addition to giving vaccinations and triaging hospitalizations \cite{abraham2022costing}. However, too many of the current visual approaches emphasize case trends, such as location-based transmission analysis \cite{antweiler2022uncovering} or studies on human mobility \cite{bao2022covid}. Few tools concentrate on the clinics themselves, which ignores the impact that clinics have had in preventing the spread of COVID-19. These clinic features (which are distinguished from clinical features) include information like the clinic’s location, business hours, the medical services provided, and their testing capacities. Clinical features are primarily based on the patient and include information such as sign levels and mortality \cite{yadaw2020clinical}.

\textit{"Testing is our window onto the pandemic and how it is spreading. Without testing, we have no way of understanding the pandemic."} Quoted statement from Ritchie et al.\cite{ritchie2020coronavirus}. As we dig deeper into the testing data, it is clear to see that high-quality and high-quantity testing can have significant economic and social ramifications during a pandemic. For instance, during COVID-19, quick testing on large scale was an essential part of the pandemic management strategy for many countries. Through these programs, cases were diagnosed and treated quickly, and the close contacts of positive cases were isolated to slow down the spread. Mass testing, isolation, and population screening in high-risk areas were all crucial public health measures that both helped to control the outbreak and provided information to guide social restrictions \cite{macintyre2020case}. We also know that accurate and prompt testing can reduce the duration of isolation and quarantine, making it faster to get back to work and resume other activities \cite{moloney2020australian}. However, clinic testing capacities are characterized by underlying uncertainty. For instance, the demographics of the area in which the clinic is located may not be fully known, or social restrictions might alter testing numbers. However difficult, it is crucial to investigate and maximize clinic testing capacities in the face of these uncertainties if we are to contain the COVID-19 outbreak.

Accordingly, we developed \textit{ClinicLens}, a visual analytics system that allows domain experts to investigate, forecast, and optimize the testing capacities of clinics. At present, \textit{ClinicLens} has been driven by an aggregated dataset of COVID-19 data pertaining to New South Wales (NSW), Australia. However, we have plans to expand the system to support other states and territories in Australia. We also note that the framework \textit{ClinicLens} is built upon could be used to provide visual analytics for other infectious diseases with a few modifications. The architecture comprises Back-end Engine and Front-end Visualization that not only provides dynamic overviews of COVID-19 trends in NSW but also includes regression models that can forecast testing capacities. Notably, the framework is robust to uncertainty. Overall, the system helps domain experts to explore the features that may affect clinic's testing capacities through a visual interface that provides analytics from multiple perspectives: spatiotemporal, location-based demographics, interventions, service factors, etc. Particularly, with the AI4VIS-oriented interactions, domain experts can investigate feature details across collaborative visual views in steps, analyze the cascading relationships between features as they develop, and iteratively update features as a way to optimize the testing capacity of clinics. Three case studies and expert interviews validate the usefulness of \textit{ClinicLens} as contributing new angles through which to explore both the trends associated with COVID-19 as well as optimize the testing capacities of clinics. In sum, \textit{ClinicLens} offers the following contributions. 

\begin{itemize}
\item \textit{ClinicLens} assesses and forecasts clinic testing capacities despite uncertainty in the data through feature modeling methods inside the Back-end Engine. 

\item This visual analytics system amalgamates the Back-end Engine with Front-end Visualization, enabling users to interactively explore COVID-19 trends across NSW and optimize the testing capacities of clinics accordingly. 

\item Three real-life case studies with expert interviews cross-evaluate the usefulness and effectiveness of  \textit{ClinicLens}.
\end{itemize}

\section{Related Work}

\subsection{Visual Analytics in COVID-19}
In a 2021 survey \cite{zhang2021mapping}, the authors built a conceptual framework derived from 668 non-academic visualizations of COVID-19 data, which summarizes who uses what data to communicate what messages in what form and under what circumstances through COVID-19 crisis visualization. From their corpus, they found that the aim of many of the visualizations (\textit{231} out of \textit{668}), i.e., \textit{34.6\%} was to communicate the severity of the pandemic to the public. For example, Paige et al. \cite{jarreau2021covid} sought to raise public awareness of this health crisis through a narrative visualization. Stepping forward to 2022, an increasing number of visual representations began to appear that were based on a more thorough understanding of the pandemic. Zhao et al., for example, introduced six main data visualizations that are frequently collaborated as public platforms in China during the COVID-19 pandemic \cite{zhao2022visualization}, while Dong et al. \cite{yu2022user} presented UcVE as a progressive spatiotemporal comparison of the COVID-19 situation in multiple locations. Additionally, Phyloview \cite{le2022phyloview}, a visual analytics system, examines the spatiotemporal evolution of COVID-19 over time, while EpiMob \cite{yang2022epimob}, a visual interface for city-wide human mobility restrictions to control epidemics, simulates changes in human mobility and the infection states that may result from implementing a particular restriction policy (or combination of policies). Olaizola et al. \cite{olaizola2022visual} also proposed a visual analytics platform designed to centralize digital contact tracing for COVID-19. However by far, the majority of these visualizations were developed to inform the public about the current state of the crisis, including any spatial and temporal restrictions put in place. Today, these visualizations are deemed to be some of the most important pieces of COVID-19 data. 

Of the remaining visualizations, \textit{21.1\%} of approaches (\textit{141} out of \textit{668}) employ a variety of other features to gauge the multifaceted impacts of the crisis for different groups. For example, some depict research, such as on public opinion \cite{han2020using,yang2022cvas},  some highlight government responses and interventions \cite{hale2021global,mathieu2021global}, and others portray environmental \cite{zyoud2021coronavirus}, economic and social \cite{praharaj2022deploying} impacts. 

A different survey, presented in \cite{tao2022survey}, includes \textit{190} references that borrow from the disaster management cycle concept \cite{li2017data} and leverage it to separate pandemic management into five phases as a loop: a pandemic outbreak, response, recovery, mitigation, and preparation for the next loop. The fact that clinics need to participate in all phases is sufficient to demonstrate the importance of clinics in pandemic management.  

Getting into the post-COVID era, the focus point needs to be gradually moved to how one should prepare for future pandemic outbreaks and their prevention \cite{leach2021post,munblit2022studying}. However, only a few visual analytics approaches provide assistance to clinics and what adjustments they should make for the future, despite the fact that clinics have played a vital role in stemming pandemics in the past. Thus, the goal of our research is to create a visual analytics system that: (1) primarily focuses on exploring and optimizing the strategies clinics use to build their testing capacities, noting that any tool developed must be robust to uncertainty; and (2) analyzes visual content from different perspectives to allow domain experts to explore COVID-19-related information. 

\subsection{AI4VIS-aided Decision-making}
Many fields have benefited extensively from AI-empowered techniques and applications \cite{cao2022new}. Among all the available tools and techniques, the concept of AI4VIS \cite{wu2021ai4vis} encapsulates how AI is applied to data visualization and how AI4VIS supports decision-making across a wide range of visual applications. Reasoning, as one of seven logics based on task-oriented classification, challenges AI to interpret visualizations in a way that derives high-level information. Through feature engineering with diverse data types, researchers have applied AI reasoning to a wide range of specific analysis tasks and decision-making problems. For example, PlanningVis \cite{sun2019planningvis} integrates an automatic planning algorithm with interactive visual explorations to efficiently optimize daily production planning in the manufacturing industry. The tool also offers support for quick responses to unanticipated incidents. Tac-Simur \cite{wang2019tac}, an AI-driven visual analytics platform, simulates the processes surrounding table tennis competitions. It helps coaches to establish competition strategies. PassVizor \cite{xie2020passvizor} offers tools for thoroughly analyzing the passing dynamics in soccer games, while CohortVA \cite{zhang2022cohortva} helps historians to identify the social structures and mobilities of historical figures by analyzing group behavior. 

To evaluate the performance of AI models in visual analytics, researchers usually employ measurement by comparing their indicators among models \cite{gollapudi2016practical}, e.g., $RMSE$, $MAPE$, and $R^2$, to further choose the proper model for their reasoning tasks. Xu et al. \cite{xu2021mtseer} exploited multiple models in mTSeer to conduct 3E forecasting with multivariate time series (3E = exploration, explanation, and evaluation). In the PromotionLens platform \cite{zhang2022promotionlens}, Random Forest (RF), Extreme Gradient Boosting (XGBoost), and Multilayer Perception (MLP) win the comparison for providing a visual exploration of strategies for promoting e-commerce commodities. In this vein, RISeer \cite{chen2022riseer} visually conducts inter-regional inspections and comparisons for urban economic development using RF and XGBoost, while LEGION \cite{das2020legion} enables users to compare and choose regression models that were created either by feature engineering or by fine-tuning their hyperparameters. 

With a backdrop, vast modeling methods in COVID-19 have been systematically reviewed in a survey that comprehensively guided challenges, tasks, methods, progress, gaps, and opportunities utilized in approaches to investigating pertinent COVID-19 analysis \cite{cao2022covid}. For example, Ou et al. \cite{ou2020machine} use AI4VIS and some machine learning-based models to anticipate gas consumption under a government intervention during COVID-19. Yang et al. \cite{yang2022cvas} apply NLP to a visual analytics system called CVAS to identify key events since the outbreak and the impact of the pandemic on public sentiment. Afzal et al. \cite{afzal2020visual} created a visual analytics prototype that gives public health professionals the ability to simulate and model the spread of COVID-19 by providing county-level data on the populace, demographics, and hospitalizations. Xu et al. \cite{xu2021episemblevis} presented a web-based visual analytics tool, EPIsembleVis, for conducting a comparative visual analysis on the consistency of COVID-19 ensemble predictions under model uncertainties. Inspired by these studies, we began to work with domain experts to collect the most relevant data and extract the most salient features from those data for further visual analysis.

\section{Prior Study}
In this section, we present our prior study conducted in consultation with a panel of three domain experts. This past work includes conventional studies on the bottlenecks and gaps in COVID-19 research, a background on the available datasets, the concept of uncertainty as it applies to clinics and testing, and the needs and expectations of a useful and valuable analytics system, as specified by our experts. 

\subsection{Conventional Research Bottlenecks and Gaps}
Over the past three years, we have been working closely with three COVID-19 experts. Each has a Ph.D. degree and more than 20 years of experience in their field. Two works for the NSW Government: expert E1 focuses on data science and visualization, while E2 is a policy consultant. Expert E3 works at a clinic.

When talking about the pandemic generally, all our experts stated that it has been a difficult period for people in Australia and for humanity as a whole. They also commented that it is unfortunate that people are still getting COVID-19 and passing away today. Although the overall momentum of the epidemic has been contained, the ever-evolving virus variants still trouble people and professionals. "\textit{We all respect and value the work of front-line medical professionals, and I want to do whatever I can to help the community improve things}," said E1. "\textit{We already have made lots of research on CT diagnosis with algorithms that aid in pre-therapy, but the conventional research on COVID-19 data stands still on numbers and simple visualizations}," he added. He also noted that he anticipates gaps and bottlenecks in the existing COVID-19 research simply because the data is multivariate and is rife with uncertainty. Existing methodologies might be helpful for analysis, but they struggle in certain difficult situations.

E1 and E2 shared with us the current visualization interface released on the NSW Government’s website. At a base level, the interface provides basic statistics for cases, tests, and deaths over the previous seven days plus overall. It also provides line charts of cases, tests, and vaccine rates as trend visualizations for the whole period. A map application keeps track of cases and test data in geo-distributions over a 30-day timeline. Nevertheless, the systems do not support any interactions of selection, filtering, or storage. Hence, the system cannot really be used to support deep decision-making. "\textit{We need intelligent decision-making aids}," E2 and E3 stated from different perspectives. "\textit{both policymakers and clinic managers need to stand by objective bases before making a decision, such as data simulation and what-if analysis, rather than acting impulsively}." "\textit{That's exactly what visual analytics can provide}." According to E1-E3’s thinking, visualization and visual analytics can greatly aid in the overall analysis of COVID-19 data. Hence, good visual analytics should be a positive step forward that should be easily combined with interactivity and machine learning algorithms to improve the overall analysis experience with specific COVID-19 tasks. It is worth noting that clinical data research makes up a sizable portion of E3’s work, and he also has experience in hospital settings. This offered us a great transformational path for connecting epidemiological data analysis to facts. He raised concerns about the entire health system being overwhelmed over the past three years – a sentiment the other experts shared. "\textit{The current healthcare system doesn't seem intelligent enough. There seems to be room for optimization because each clinic operates independently and has only a few connections to one another}," E3 emphasized. 

\subsection{Data Background and Uncertainties}
Much information has been provided by the COVID-19 Data Program \cite{NSW2022data} to improve the NSW Government’s coordinated COVID-19 response. Headed by the NSW Data Analytics Center (DAC), this program assembles open-source datasets of COVID-19-related information, such as tests, cases, and clinic information. The datasets are updated weekly and made available to the public on the NSW Government website. However, in the post-COVID era, analysis tasks have changed from straightforward numerical analysis to more intricate aftermath work. Therefore, some of the statistical items in the datasets have been modified given that the epidemic has largely been suppressed. Likely infection sources, for instance, are no longer identified, but geo-based clinic information has been added to help people locate where to obtain a test more quickly.

Our first focus as researchers was on the datasets that are still being continuously updated. These included the tests, cases, and clinic datasets from the NSW Government website. E2 pointed out that government interventions affecting COVID-19 restrictions should also be taken into account: "\textit{Overlapped interventions should have significant impacts on test and case amounts}". Hence, we also crawled news on government interventions, which we combined with the case information on confirmed infections, the test results, their locations, and the notification dates. Aside, the clinics' dataset contained details of the COVID-19 testing and assessment centers, such as their geographic location, services list, and business hours. Although these three datasets appear straightforward and manageable, all domain experts still noted that they are actually complex and filled with uncertainty. 

In clarifying these uncertainties, we discovered three main areas of issue: 1) the method used to count the tests; 2) daily test attributions; and 3) each clinic’s daily testing numbers. 

In terms of the counting methods, our domain experts noted that \textit{"The NSW Government has a specific counting method on tests."} The NSW Government determines counting every negative COVID-19 test on any day separately (i.e. \includegraphics[scale=0.06]{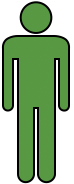} $\Rightarrow n \cdot $\includegraphics[scale=0.06]{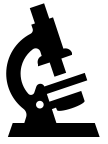} = n). That means an individual’s first positive test is counted, along with each negative test they’ve had on previous days. But, after the first positive test, no additional tests are included (i.e. \includegraphics[scale=0.06]{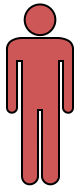} $\Rightarrow n \cdot$ \includegraphics[scale=0.06]{figures/symbols/Testing.png} = 0 ). All three experts acknowledged that this form of counting guarantees the authenticity of the tests to at least some extent. Plus, it reduces errors and makes statistical analysis easier. 

Second, \textit{"We must realize that not all daily tests summed each day are from the previous day."} The truth is that not all clinics have the ability to test COVID-19 nucleic acid results, and shipping every test from clinics to a specific laboratory for testing delays the release of data. We were reminded that, usually, all the test results would be made public within 3 days, although some may still take up to seven days. Thus, the second uncertainty is that the actual daily tests released are dispersed but should be reported in the data releases of the previous days. Hence, as researchers, we needed to consider how to handle this uncertain consequence when modeling clinic features.

Third, our experts reported that \textit{"we cannot monitor the number of tests or testing capacities provided by each clinic on a daily basis."} because the basic unit under which the testing numbers are released is either a local government area (LGA) or postcode, not each individual clinic, which also leads to uncertainty. 

On the whole, the first uncertainty, counting methods, can be managed in the data processing. To resolve the last two uncertainties, dispersed daily test counts and the lack of attribution to particular clinics, the domain experts concurred that the testing capacities for each clinic should be abstracted by dividing them into daily test numbers for each clinic. Then machine learning algorithms could be used to establish models for predicting the daily test count amounts allocating this uncertainty.

\subsection{Requirement Analysis}
We finalized the brainstorming with E1-E3 by gathering their concerns and expectations and summarizing their requirements. We had all settled on visual analytics as a significant theme, and all experts agreed that they expected to receive a visual analytics system that used the continuously updated NSW COVID-19 data. After several rounds of discussion, we concentrated on the following four requirements. 

\textbf{R1. Establish feature modeling for clinic testing capacities given uncertainty.} Our approach should first collect and sort the existing data. This would involve untangling the selected features to make them reasonable given the uncertainty in the data. Next, each data feature should be ported into an aggregated dataset. Any data scaling that may affect a clinic’s testing capacities would be formulated, and feature models would be established through feature engineering. Last, we would undertake an evaluation to confirm the model’s reliability. To accomplish all this, it would be necessary to implement Back-end Engine that included all the above processes as components. 

\textbf{R2. Visualize COVID-19 information and clinic-based test capacities based on multiple features.} On the basis of our prior study, the domain experts expected that we would implement Front-end Visualization that could display the time series and location-based trends found in the data, as this is what was required to support detailed and dynamic visualizations of daily test and positive case numbers. Additionally, machine learning algorithms and visual analytics techniques should be used to provide multiple different user-friendly views as the system’s outputs. As the foundation for visualizing clinic-based information in the COVID-19 background with numerous features, these views need to cooperatively support exploration navigation and intuitive interactions. 

\textbf{R3. Provide the ability to investigate the impact of possible features on the clinic's testing capacity.} The experts also asked for the flexibility to interact with any of the features that could affect testing capacity. "Advanced interacting" might include inspecting each feature’s importance, filtering the time periods, selecting the locations to be included, adjusting the clinic features included in the analysis, and saving the results of exploration. They also wanted a system that allowed iterative investigations of these impacts and one that, in addition, offered a well-designed visual view of the ground truths, initial prediction results, and updated prediction results in any given analysis. 

\textbf{R4. Enable strategies for optimizing a clinic’s testing capacities.} Above all, the experts expected a comprehensive system where the knowledge gained from their iterative investigations could be used to optimize clinics' testing capacity and inform upcoming decisions. Adopting reasonable strategies requires the application of case studies in real-world scenarios to enhance the resilience of clinic testing capacity for sustainable balances and to demonstrate the effectiveness of our system.

\begin{figure*}[!t]
\centering
\includegraphics[width=7in]{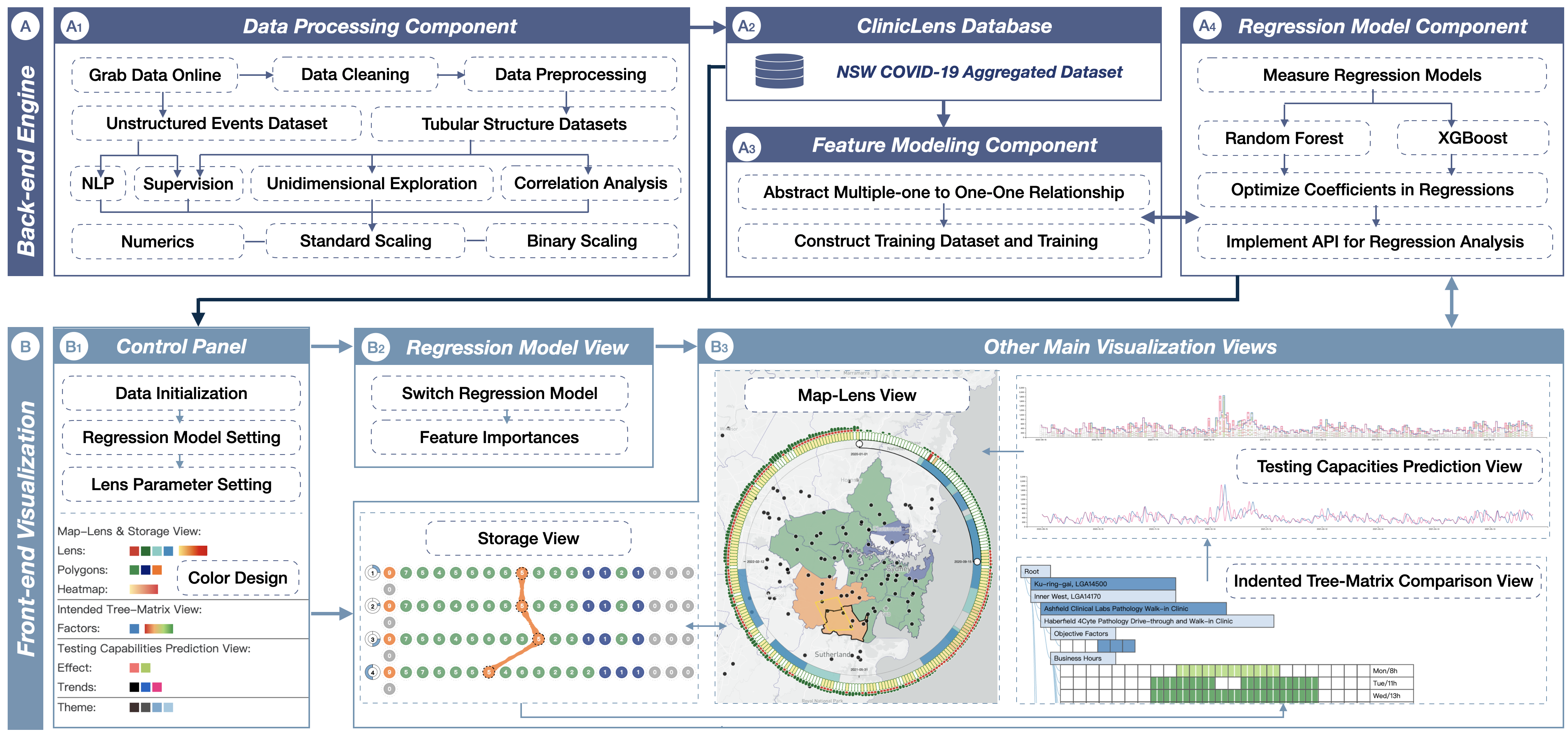}
\caption{The \textit{ClinicLens} framework. \textit{ClinicLens} consists of (A) Back-end Engine and (B) Front-end Visualization. The entire pipeline of the framework is depicted here, which begins with importing the data in A1 via the \textit{Data Processing Component}. The data processing procedure involves multiple steps but results in an aggregated NSW COVID-19 dataset that contains three different feature types. These data are then stored in a database (A2). In A3, the \textit{Feature Modeling Component} loads the aggregated dataset, abstracts one-to-one relationships, and constructs the training set. The \textit{Regression Model Component} then establishes two appropriate models (RF and XGBoost) for data prediction in A4. The entire Back-end Engine is always ready to be called upon by the Front-end Visualization system (B). A suite of parameters can be adjusted in the Control Panel (B1). Then users are transported to the regression model view (B2). Additionally, four other main visualizations can be rendered by the front-end system for further interpretation, exploration, and interaction (B3).} 
\label{fig:Overview}
\end{figure*}

\section{ClinicLens Overview}
ClinicLens was developed to meet the above four requirements, which is an interactive visual analytics system based on regression analysis that was designed to explore and optimize clinic testing capacities using data that contains uncertainty. As outlined in Fig. \ref{fig:Overview}, the \textit{ClinicLens} pipeline is based on the experts’ requirements. It consists of Back-end Engine and Front-end Visualization. The Back-end Engine is composed of \textit{Data Processing Component}, \textit{Feature Modeling Component}, and \textit{Regression Model Component}. The regression model quantifies the data columns to predict the daily testing capacity of each clinic as a fundamental implementation of R1. As the assembled dataset driving \textit{ClinicLens}, we combined the raw COVID-19 test and case information published by the NSW Government with the clinic data, data on the government interventions, and the demographic census data. These datasets were then cleaned, extracted, aligned, and rescaled into usable features that could be used to train the regression models. The Front-end Visualization conjuncts multiple views guided by elaborate colors \cite{shi2022colorcook}. Users can interact with these views to explore trends in COVID-19 tests and cases based on location (to meet R2) or clinic and clinic features (to meet R3). Additionally, the system can generate iterative forecasts of testing capacity based on any of the features available in the system (to meet R4).

\section{Back-end Engine}
The Back-end Engine in \textit{ClinicLens} involves an exhaustive data pathway, from accepting the raw data to processing and storing that data in database to finalizing the output of the two trained models. Notably, the Back-end Engine is robust to data uncertainty. The models, based on random forest (RF) and XGboost, are also bridged to the Front-end Visualization to both display the data in a way that can be iteratively analyzed and to offer predictions of clinic testing capacities. The overarching goal of the framework is to predict each clinic’s testing capacity using the daily LGA testing numbers as a ground truth. Additionally, each component of the Back-end Engine connects to the front-end system via an API port. Each of the components is described in more detail next.

\subsection{Data Processing Component}
The purpose of the \textit{Data Processing Component} is to load the raw datasets retrieved from Data.NSW, NSW Health, and the Australian Bureau of Statistics (ABS) into the framework. These datasets comprise both structured and unstructured data. In terms of structured data, three datasets contain tabular data. 1) The NSW government COVID-19 tests and cases data, which include the number of daily COVID-19 tests conducted \footnote{NSW COVID-19 tests: https://data.nsw.gov.au/nsw-covid-19-data/tests} and the number of confirmed cases \footnote{NSW COVID-19 cases: https://data.nsw.gov.au/nsw-covid-19-data/cases}, are released on a weekly basis. These data are classified by LGA and postcode. 2) COVID-19 clinic data. This dataset contains the service details of the authorized clinics \footnote{COVID-19 clinics: https://data.nsw.gov.au/nsw-covid-19-data/covid-19-clinics}, hereafter termed the "clinic features". These details describe such things as each clinic’s business hours, services offered, and testing requirements. Again, these data are classified by LGA and postcode. 3) The 2021 Census dataset, which contains location-based demographic information gathered from the most recent census \footnote{The 2021 Census: https://www.abs.gov.au/census}; the unstructured data pertain to government interventions. These data were crawled and then processed using a natural language processing (NLP) model as part of prior research \cite{yu2022user}.

To prepare \textit{ClinicLens}, each dataset was next pre-processed to remove all self-testing counts and self-reported cases from the test and case numbers. In tandem with confirming that there were no empty values in any of the datasets, this process aligned the data with the case counts released by the clinics. The COVID-19 tests and case datasets were then combined with the clinic features and the government interventions, yielding an aggregated dataset of \textit{252,350} rows. These data were stored in a database spanning \textit{128} LGAs (\textit{612} postcodes) and \textit{248} clinics and \textit{10} government interventions over a \textit{1030-day} period from 1 January 2020 to 28 October 2022.  

Before feature modeling, the aggregated dataset was aligned and rescaled by feature. We were guided by the domain experts, optimizing and selecting the features by assembling and recalculating them in a series of iterative experiments. Through unidimensional exploration and correlation analysis, we ensured that every feature was both reasonable and distinguishable. Thus, our assumptions about which features could affect a clinic’s testing capacity were built up over time and guided by expert input. Overall, these features fall into three main categories. 1) Numeric features ($ {X_N} $), such as LGA population densities, daily test, and case numbers, the clinics’ business and break hours, location-based clinic counts, and day counts over the whole period. 2) Standard scaling features ($ {X_S} $), which consist of the current day of the week (1-7), the current season (1-4 represent Spring to Winter), and three levels of government interventions (0-3, where 0 means no intervention and 3 means the strictest interventions). 3) Objective factors of clinic features ($ {X_B}_m $), mostly binary, which comprise factors such as "Referral Required" (0/1 indicating whether a GP referral was required to conduct the test), "Age Limit" (whether the clinic tested infants), "Booking Required" (whether an online booking was required before testing), "Walk-in Allowed" and "Drive-through Allowed", and "Wheelchair Accessible", which are all self-explanatory.

\subsection{Feature Modeling Component}
The \textit{Feature Modeling Component} begins with abstracting the aggregated dataset and preparing the training dataset, all for the purpose of predicting the testing capacities of each clinic. To illustrate the process as simple as possible, we used LGAs as our basic unit; however, postcodes could also be used as an alternative unit.  

Compared to the one-clinic-one-LGA relationship (one-to-one), a multiple-clinics-to-one-LGA relationship (multiple-to-one) will introduce uncertainty into the clinics’ daily test counts. As previously discussed, we did not have information on the daily test counts of each clinic to directly support multiple-to-one relations, so we redefined a new one-to-one relationship for each clinic in each multiple-to-one relationship for the purposes of training. Specifically, we reconstructed a new one-one relationship for each multiple-one relationship to contain all clinics in a new LGA-based entity by rescaling their clinic features to LGA-based wholes so that they could correspond to the daily released test amounts on each LGA as ground truth. Hence, each clinic binary scaling feature and clinic count in multiple-one are rescaled, to sum up as a whole feature: $X_{B(m,n)}=\{{\textstyle \sum_{0}^{n}} x_1,{\textstyle \sum_{0}^{n}} x_2,\ldots,{\textstyle \sum_{0}^{n}} x_m, n\}$, where $ m $ represents the number of binary factors and $ n $ stands for the number of clinics in this multiple-one relation. Other features are automatically transcribed to the previous value based on LGAs, which aligns the ground truth of daily test amounts with the LGAs of a single clinic. Thus, the assumption of aiming regression training in each LGA is defined as $ Y_{LGA}=\{{X_N},{X_S},X_{B(m,n)}\}$.

\begin{table}[tb]
   \caption{Regression Model Performance.}
  \label{tab:model}
  \scriptsize%
	\centering%
  \begin{tabu}{
	l%
	*{3}{c}%
	}
  \toprule
   Models & $RMSE$ & $MAPE$ & $R^2$   \\
  \midrule
	Linear & 770.65 & 141.86  & 0.27 \\ 
        GBDT & 135.20 & 81.50 & 0.97  \\ 
        \textbf{XGBoost} & \textbf{69.90} & \textbf{66.61}  & \textbf{0.99}  \\ 
        CatBoost & 197.16 & 85.95  & 0.95  \\ 
        ExtraTree & 379.57 & 73.16  & 0.82  \\ 
        LightGBM & 184.45 & 79.88  & 0.96  \\ 
        DecsionTree & 337.89 & 73.98  & 0.86 \\ 
        \textbf{RandomForest} & \textbf{59.62} & \textbf{70.23}  & \textbf{0.99}  \\
   
  \bottomrule
  \end{tabu}%
\end{table}

\subsection{Regression Model Component}

As COVID data involve time-series information, regression models serve as the common choices for forecasting the testing capacities \cite{cao2022covid}. We thus tested various models, including linear regressor, GBDT, XGBoost, CatBoost, ExtraTree, LightGBM, Decision Tree, and RF on our constructed training dataset. The results are shown in Table \ref{tab:model}, which are evaluated by $RMSE$, $MAPE$, and $R^2$. Of all models tested, the comparison result suggests that both RF and XGBoost are the most appropriate regression models for this analysis task because they have representative performance on predicted accuracy and feature importance. We determined to employ these two regression models and applied them to the aggregated dataset to predict the possible testing amount for each clinic per day $y_{clinic}$. The regression estimates the daily clinic test by a regressor $f$ like: $ y_{clinic}= f(\{{x_N},{x_S},x_{B(m,1)}\})$. $y_{clinic}$ is evaluated per the ground truth of $Y_{LGA}$, as $Y_{LGA} \approx {\textstyle \sum_{1}^{n}y_{clinic}} $. We adjusted and optimized the coefficients to ensure the outputs are reasonable, such as ensuring the test amounts not to be negative and keeping two decimal digits.

\section{Front-end Visualization}
We introduce the Front-end Visualization of \textit{ClinicLens}, which enables iterative closed-loop explorations of clinic testing capacity across multiple collaborative visual views. The in-process visual exploration is followed by requirements and started from inspiration by the visual metaphor "Lens on Map \cite{ma2020gtmaplens}", concurrently implementing other views and interactions that dynamically bridge the connection to the Back-end Engine. 

In addition to the four main visualization views, there are two supplemental views: the Control Panel and the Model Features View. The Control Panel allows users to select particular datasets, filter timelines, nominate which unit to use, select a preset set of features or all features to include in the regression, adjust the lens size, and change the color representations. The Model Features View shows the RF and XGBoost switches, along with the importance attached to each feature to help determine which model is most appropriate for the user.  

Turning to the four main views, the \textit{Map-Lens View} offers options for navigating, inspecting trends in the pandemic, selecting LGAs, and a heatmap of the clinics’ testing capacities given the current settings. The Storage View is where users can save their exploration results. The results of explorations can be saved at each iteration in the form of ranking or tracking LGAs by their testing capacities. The \textit{Indented Tree-Matrix Comparison View} is designed to help users interactively inspect, compare, and configure the clinic features to generate testing capacity predictions, while the \textit{Testing Capacities Prediction View} depicts the ground truth trends before and after any predictions are made. More detail on each of these views is offered next. 

\subsection{Map-Lens View}
\begin{figure}[h]
\centering
\includegraphics[width=3.5in]{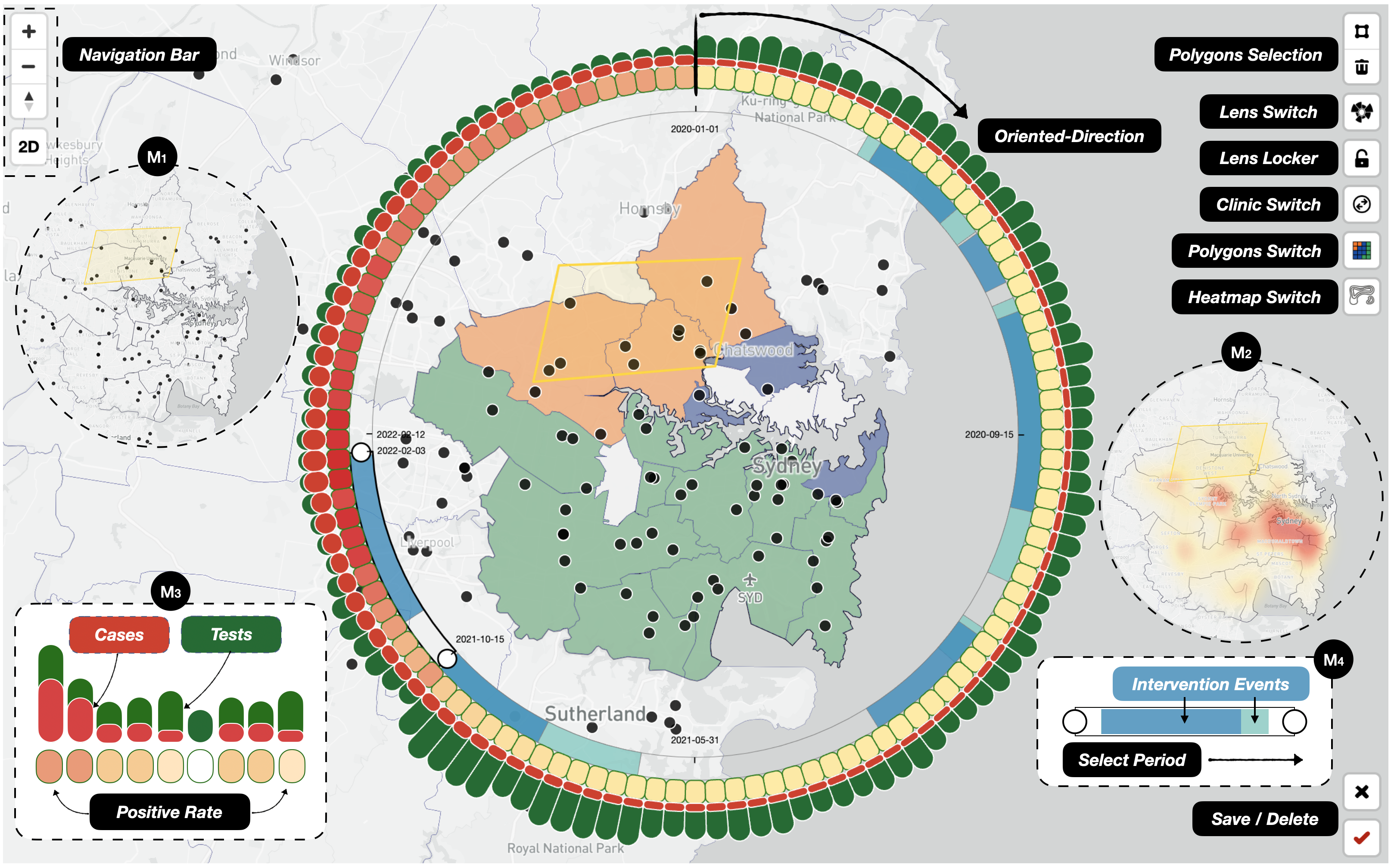}
\caption{The Map-Lens comprises three nested layers: (1) The intervention timeline layer (M4), which encodes interventions and is adjustable through draggable sliders; (2) The test and case count layer (M3), which are displayed through color bars; and (3) the positive case rate layer (M3), rendered as a heatmap. The four buttons (Lens Switch, Clinic Switch, Polygons Switch, and Heatmap Switch) control the lens and the three superimposable map layers. These three layers are the base layer (M1), the status polygons layer, and the heatmap layer (M2). The lens can also be locked in place on the active scope using the Lens Locker button. When finished exploring the \textit{Map-Lens View}, users can preserve the results with the Save button or Delete the callback from the Storage View.}
\label{fig:Lens}
\end{figure}

The \textit{Map-Lens View} is designed to serve as a trigger that instructs users to begin their exploration. As such, this view offers a variety of buttons that allow interactivity, as illustrated in Fig. \ref{fig:Lens}. A well-designed lens, consisting of three nested layers, is attached to the map’s scope once a size parameter has been selected in the Control Panel. We specify these three nested layers to proceed clockwise, starting from the vertical and stopping at the circular end.  

\textbf{Inner: The intervention timeline layer.} The inner layer assembles the features of the time series and the interventions over the selected period, including the level of intervention from 0-3. A two-sided draggable slider is available, which can be used to filter the timeline by day, and two colors (turquoise blue \includegraphics[scale=0.5]{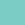} for eased events and saxe blue \includegraphics[scale=0.5]{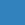} for restrictions) to distinguish 0-3 level-divided intervention events, where null means no event and the superimposed two colors represent the strictest intervention. 

\textbf{Outer: The test and case count layer.} The outer layer expands the space to evenly distribute and radially express the daily test and case counts based on the chosen time frame. To balance space utilization and data range, the radial height $V$ of both test and case amounts $H$ is considered to combine piece functions that are equal to the same calculation: $H_{(day,tests,cases)} \propto a \cdot \ln_{}{V(tests,cases)} + \frac{V(tests,cases)}{b} $. $V(tests,cases)$ means the amount number of either cases or tests, $a$ and $b$ are adjustable parameters used to achieve a better screen fit. Different representations were painted, where true green \includegraphics[scale=0.5]{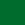} stands for tests and true red \includegraphics[scale=0.5]{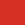} represents the cases.

\textbf{Middle: The positive case rate layer.} The daily rate of positive cases helps users to better understand the relationship between testing capacities and cases. It is essentially a highlighted color scale (from canary yellow to scarlet red \includegraphics[scale=0.5]{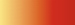}) to map the daily COVID-19 severity of LGAs in the current Lens scope. The positive case rate for any given day is calculated by $R_{day} = \frac{V_{cases}}{V_{tests}}$.

In addition to basic navigation tools, the \textit{Map-Lens View} likewise offers some buttons for users to interact with the map. Users can draw polygons to select LGAs or use nearby buttons to toggle each map layer on or off. The map’s scope is made up of three superimposable layers: a base layer with pins representing clinic locations, a status polygons layer, and a heatmap layer of testing capacities. The status polygons and the heatmap of average testing capacities interact simultaneously with lens pans and zooms of the map. 

The map base layer is used for locating each clinic in NSW. Each clinic is included in an LGA. The basic setting of the status polygons layer is based on the number of clinics in an LGA. To make it simpler for users to visually recognize multiple-one or one-one relationships, we encoded the sea green \includegraphics[scale=0.5]{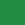} for LGAs with multiple clinics and the wisteria purple \includegraphics[scale=0.5]{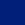} for an LGA with a single clinic. In addition, we commit to utilizing tangerine yellow \includegraphics[scale=0.5]{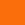} for user-selected LGAs through interactions. The testing capacities layer shows the average testing capacities of each clinic on a heatmap scale from moon yellow to strong red \includegraphics[scale=0.5]{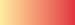}. The average testing capacity of a specific clinic is calculated by: $T_{clinic} =  \frac{{\textstyle \sum_{d_1}^{d_2}} y_{clinic}}{d_2-d_1}$.

\subsection{Storage View}
User-centered design is being incorporated into an increasing number of visual analytics systems. In \textit{ClinicLens}, we employ a user-centered design that is similar to our previous design in \textit{UcVE} \cite{yu2022user} to support the ability to save the results of users’ explorations for later recall. The \textit{Storage View} receives the save command and further compresses and lists the LGAs that \textit{Map-Lens View} had selected for sequential allocation, according to the exploration results on \textit{Map-Lens View}. With the exception of the first circle in each sequence, which is called the location-unit circle and shows the sequence number and the selected period in \textit{Map-Lens View}, each LGA queried in the sequence is represented by an LGA circle with the same colors as the LGA on the map. The precise clinic numbers in each LGA, represented by the number-in-center LGA circles, are queued in descending order based on the total number of tests in the sequence.

As depicted in Fig. \ref{fig:Overview}, the \textit{Storage View} additionally allows for the reordering, tracking, and highlighting of every sequence with associated links if they point to the same LGA across different sequences (e.g., LGA Canterbury-Bankstown is tracked rankings by link connections among sequences). The situations of all LGA circles in each historical or current storage sequence can also be updated on the \textit{Map-Lens View} as callbacks.

\subsection{Indented Tree-Matrix Comparison View}
\begin{figure}[h]
\centering
\includegraphics[width=3.5in]{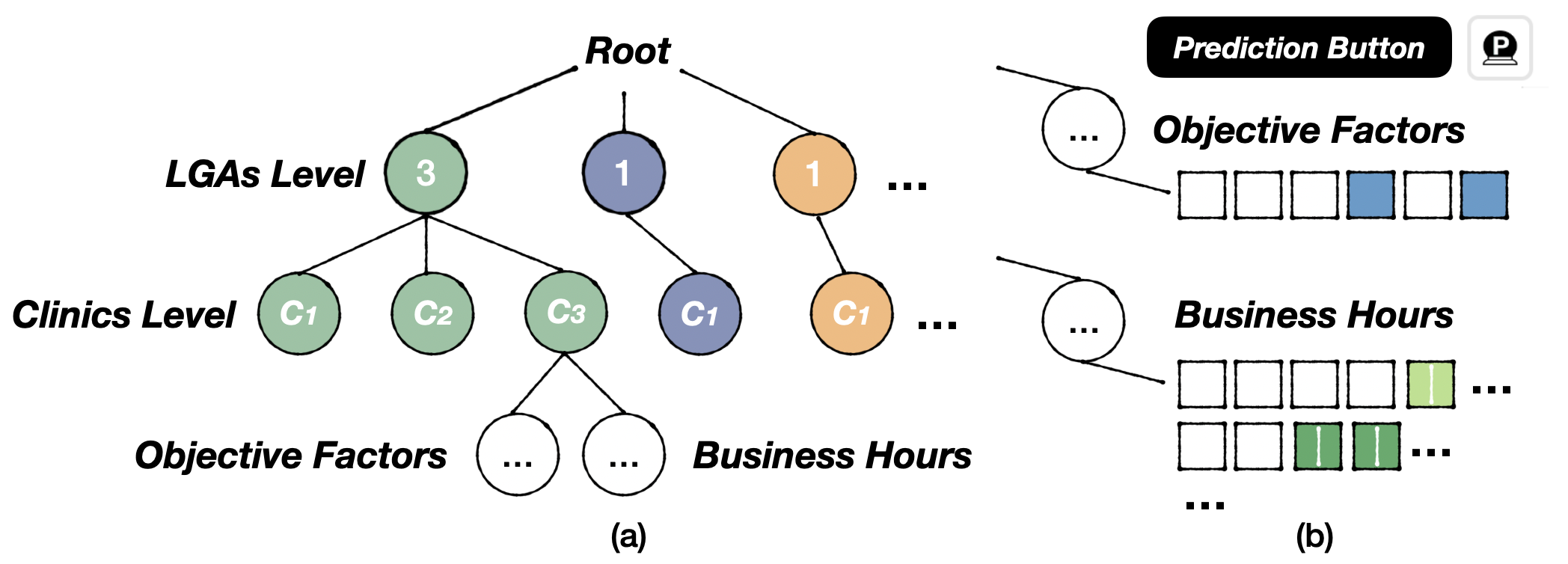}
\caption{The \textit{Indented Tree-Matrix Comparison View} combines the benefits of a tree structure and matrices to interactively expand the hierarchy and balance any adjustments to the objective factors and the business hours. Here, (a) shows the entire structure of the Tree-Matrix View, while (b) illustrates how a unified vector and matrix represent both the features of the objective factors and the business hours.}
\label{fig:Indented}
\end{figure}

Inspired by the hierarchical confusion matrix in Neo \cite{gortler2022neo}, we created and implemented an indented tree-matrix structure of the clinics’ features by LGA. Abstracting a tree structure, as shown in Fig.\ref{fig:Indented}, the hierarchy progressively represents LGAs, clinics, and features by level. 

The matrix on each leaf node shows the specific features of the clinics, while the hierarchical structure in Fig.\ref{fig:Indented}(a) illustrates the progressive relationship between the LGAs, clinics, and features. Subtrees can be expanded or collapsed, and features can be revised by clicking. For a more comprehensive analysis, users can add LGAs to the tree-matrix by clicking the LGA circles in the Storage View.  

Additionally, each clinic’s leaf-fused features can be split into two blocks, as shown in Fig.\ref{fig:Indented}(b). One block is a 1 × 6 vector of blocks representing the clinics' objective factors. (From left to right, these are Referral Required, Age Limit, Booking Required, Walk-in Allowed, Drive-through Allowed, and Wheelchair Accessible). Salvia blue \includegraphics[scale=0.5]{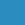} is used to distinguish whether the (binary) value is "Yes" or "No"). The other block is a 7×48 matrix of blocks detailing the business hours across a week from 0:00 to 24:00 for each day of the week from Monday to Sunday. Here, a scale of red to green indicates from few to many hours, with each block in the matrix representing half an hour as the basic unit to allow for precise adjustments. Users can compare and edit the various features of the adjacent vectors or matrices among the clinics using the \textit{Indented Tree-Matrix Comparison View}. The prediction button queries the Back-end Engine for updated features and performs regression analyses of the clinics’ testing capacities. 

\begin{figure*}[!t]
\centering
\includegraphics[width=7in]{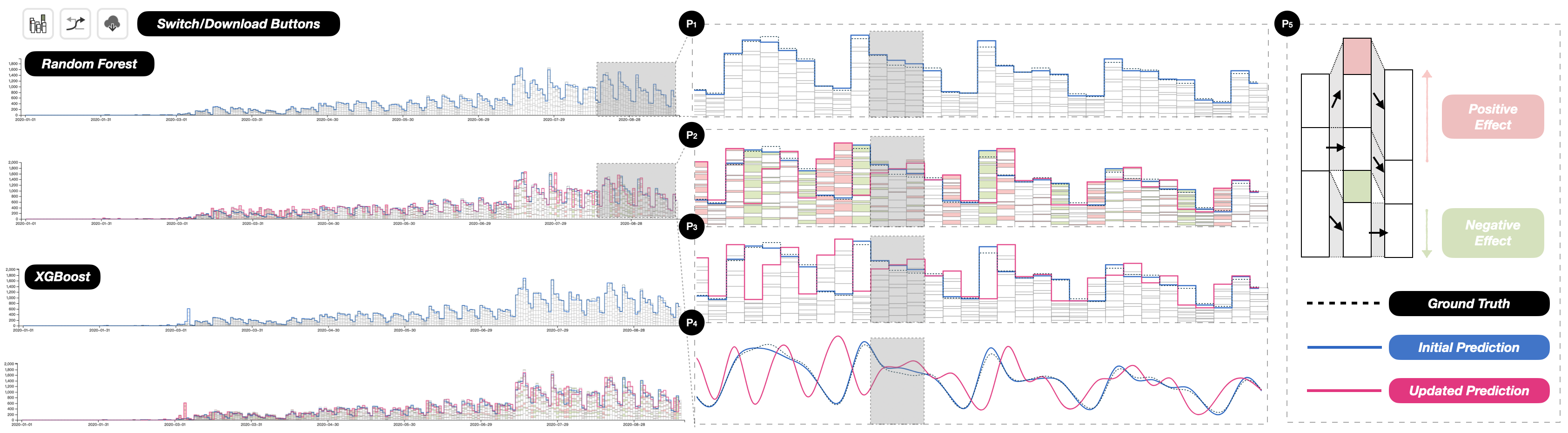}
\caption{The \textit{Testing Capacities Prediction View} provides two alternative regression models, RF and XGBoost, which are each based on different feature importances. The four statuses (P1-P4) are based on different data inspection angles and are controlled by corresponding buttons. These can be summarized into two modes: (P1-P3) Step-line mode and (P4) Curve-line mode. (P5) shows the five colors, which designate: positive and negative effects, the ground truths, the initial predictions, and the updated predictions for clinic testing capacities by LGA.} 
\label{fig:Prediction}
\end{figure*}

\subsection{Testing Capacities Prediction View}
The \textit{Testing Capacities Prediction View} relies on two regression models. These models provide users with insights to help interpret specific trends and correlation variations in the testing capacities of each clinic in a given LGA over the chosen time period. This view offers two modes of inspection across four views, delineated by five colors, as shown in Fig.\ref{fig:Prediction}. The prediction results can be saved as a graph. 

\textbf{The two modes of inspection:} 1) a step-line chart that conveys the ground truth against the predicted daily testing capacity for each clinic (along with its possible variations); and 2) a curve-line chart that shows smooth trends and discrepancies between the ground truths and the predictions. 

\textbf{The four switchable views:} Users can switch between the four views via buttons with each offering a different data inspection angle. The first view, the \textit{Testing Capacities Prediction View}, shows the predicted test counts for each clinic as bars under a daily total cap versus actual testing capacities (check P1 of Fig.\ref{fig:Prediction}). The Second is a view of the updated predictions. This view displays the forecast test capacities for each clinic after revisions to their features have been made. Total capacities are shown as step-line charts for negative effects, and these are connected to step-line charts of the original predictions and the ground truth (see P2). The third view shows the updated predictions of the test counts for each clinic based on any changes made to the features. This view also includes step-line charts of the original predictions and the ground truths without positive and negative effect bars, as shown in P3. Last are the three curve-line charts representing the ground truth, the initial predictions, and the updated predictions, as depicted in P4. 

\textbf{The five colors:} These colors need to be reasonable and distinguished by other colors used, we borrow true black \includegraphics[scale=0.5]{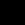} the dashed lines representing the ground truth trends; ultramarine blue \includegraphics[scale=0.5]{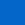} for the solid lines representing the initial predictions; fuchsia red \includegraphics[scale=0.5]{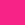} for the solid lines representing the updated predictions; fuchsia red \includegraphics[scale=0.5]{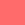} for the positive effects of any changes made to the clinic features; and lime green \includegraphics[scale=0.5]{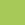} for the negative effects of any changes made to the features.

% \section{Implementation and Deployment}
% We engaged the Back-end Engine, driven by Node.js \cite{tilkov2010node} and Python libraries (Flask, Numpy, and Sklearn), to store datasets in a MySQL database, and the Front-end Visualization, implemented on Vue.js and supported by D3.js \cite{bostock2011d3} and Mapbox.js, to develop  \textit{ClinicLens}. The entire approach has been open-sourced in the GitHub repository  (\textit{https://github.com/YuDong5018/clinic-lens}) and includes all used datasets, the training models, regression analysis algorithms, and visual analytics system. By following configuration and guidelines,  \textit{ClinicLens} is simple to deploy on a server or localhost address. 

\section{Evaluation}
To evaluate ClincLens, we conducted three case studies, after which we gathered and synthesized feedback through interviews with domain experts. 

\subsection{Case Study I: Overview of COVID-19 Broad Trends and Regional Clinic Average Testing Capacities}
\begin{figure}[h]
\centering
\includegraphics[width=3.5in]{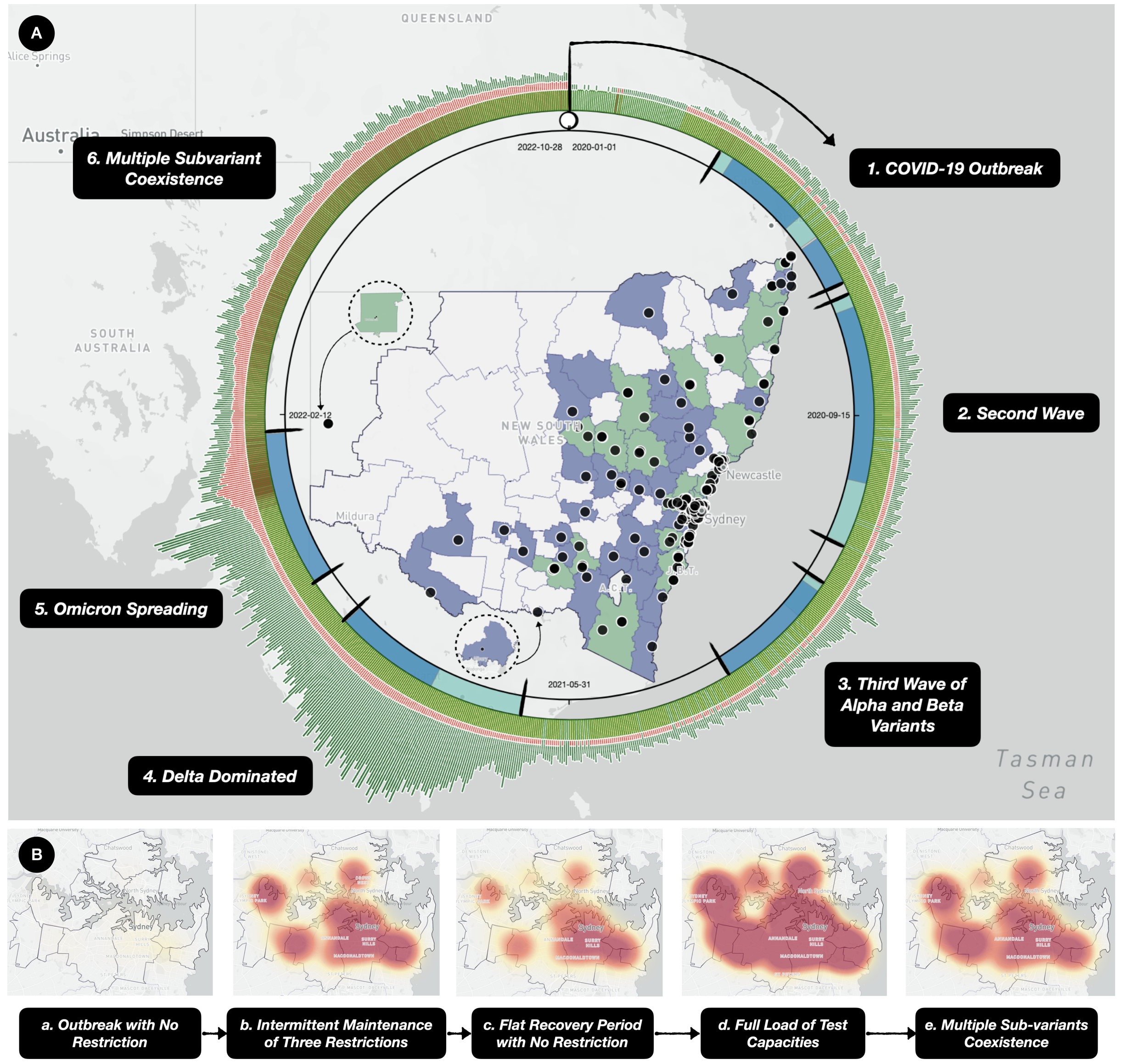}
\caption{Case Study I: (A) The COVID-19 broad trends in NSW from 1 January 2020 to 28 October 2022, where intervention events in the timeline represented government intervention events against the virus variants in varied periods and the scatter points represented the varied distribution of clinics. (B) A five-piecewise periodic heatmap chain of the average testing capacities of clinics in the Greater Sydney Area.}
\label{fig:Lenscase1}
\end{figure}

We first conducted a general exploration of broad COVID-19 trends across all of NSW using \textit{ClinicLens}. After initializing the parameters to include the entire period from 1 Jan 2020 to 28 Oct 2022, we drove the lens to include all LGAs. The broad trends returned are shown in Fig. \ref{fig:Lenscase1}(A). 

Overall, we observed the following: 

1) Generally, the daily test counts had a periodic 7-day trend, except for two irregular patterns: a) a pulse at the beginning of the third wave of the Alpha and Beta variants; and b) a peak and a sustained high level of testing beginning in the second half of 2021. Then, from the middle of the NSW Government’s final intervention, testing counts decreased. These were the reopening restrictions put in place between 8 Nov 2021 and 31 Jan 2022. 

2) In the early days of the outbreak, daily case numbers were generally low. There were even scattered days when there were no new cases. However, the situation started deteriorating in 2022, peaking at \textit{20.9k+} cases on 6 January. The spreading of the Omicron and other multiple subvariants led to a sharp increase in case numbers, which maintained high until the end of the study period. 

3) Because logarithmic mappings of the test and case numbers are not intuitive at small values, we instead examined the daily positive case rates. First, we divided the entire time period into five segments, each with its own specific COVID-19 background. Phase 1 covered the period of the outbreak before any restrictions were put in place, this being from 1 Jan 2020 to 24 Mar 2020. Positive cases were reported on January 25 and 27, 2020, followed by an approximately one-month absence of cases before COVID-19 finally broke out in force at the beginning of March. After reaching a peak of nearly \textit{1,000} cases per day in late March, the government decided to implement a limited intervention. Phase 2 covers a period of three main restriction programs enacted between 25 Mar 2020 and 26 Feb 2021. These interventions successfully suppressed the spread, each within a month, giving rise to a 10-month period of diagnosis rates under \textit{0.2\%} against three waves of the virus. Phase 3 spans a flat recovery period with no interventions from 27 Feb 2021 to 24 Jun 2021. During this period, NSW’s clinics conducted over \textit{10,000} tests per day. Generally, positive case rates stayed under \textit{0.1\%}, except for one day on 4 Mar 2021, when it reached \textit{0.15\%}. In Phase 4, from 25 June 2021 to 31 Jan 2022, the Delta and Omicron variants successively overtook all measures to stop them. In this period, the testing capacities of all clinics were overwhelmed to full loads. After two temporary bursts of spread around the Christmas and New Year holidays, positive case rates crossed the \textit{40\%} mark and, unsurprisingly, continued to rise to a top rate of \textit{58.16\%}. By 8 Jan 2022, case numbers had reached \textit{149,033} from \textit{256,229} tests. The final phase covers the period from 1 Feb 2022 to 28 Oct 2022, where multiple subvariants coexisted in the population. During this period, the clinics’ test capacities gradually restored to levels allowing over \textit{10k} tests per day. However, given the voracity of the subvariants, positive case rates stayed high. Fortunately, none exceeded \textit{40\%}, and the majority were below \textit{25\%}. 

We then explored the clinic distribution across NSW, observing that clinics had been established based on populated areas. For example, most of the LGAs near the coast had more than one clinic with fairly concentrated clinic densities. By contrast, LGAs in the more inland areas had few authorized clinics. For example, Broken Hill, a regional service town, had only two clinics to serve the entire Far West Region – an area of around \textit{95,000} {$km^2$}. This observation gave us some insight into the huge differences in test counts between the coastal and inland regions. Hence, we used the heatmap to decompose the major trends around the Greater Sydney Area by the five identified phases of the pandemic, as shown in Fig.\ref{fig:Lenscase1}(B). These heatmap results again confirmed our findings of the broad changes in periodic testing capacities across the clinics in the Greater Sydney Area.

\subsection{Case Study II: Investigating the Impact of Clinic Features on Testing Capacities}
In this case study, we explored how the clinic features influence testing capacities. As discovered in Case Study I, most clinics were operating at full capacity from 25 Jun 2021 to 31 Jan 2022. Hence, we began our investigation by adjusting the map lens to this period. 

\begin{figure*}[!t]
\centering
\includegraphics[width=7in]{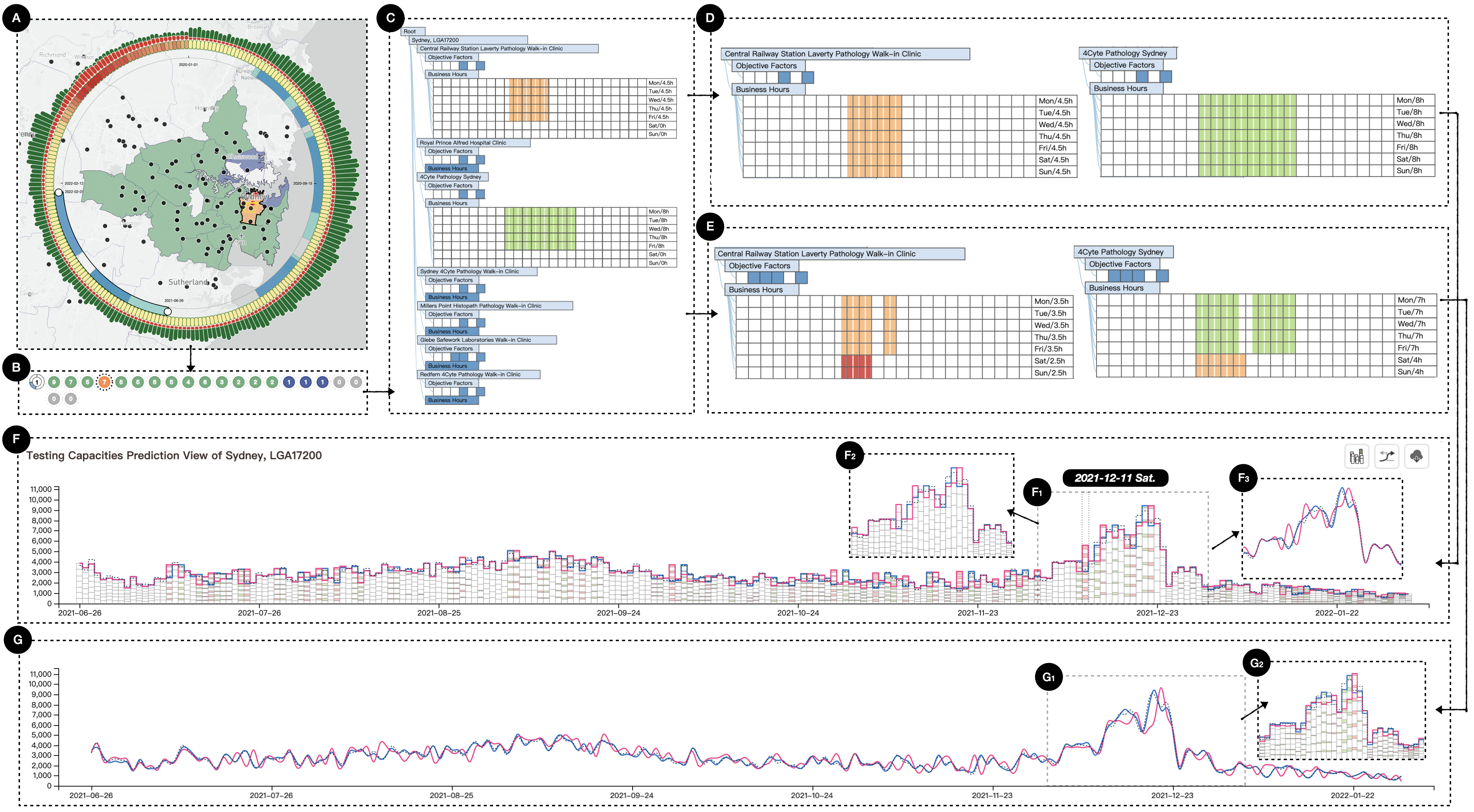}
\caption{Case Study II: (A) Driving Map-Lens to the Greater Sydney Area within a time constraint of 26 June 2021 to 1 February 2022. (B) Saving the exploration result to the Storage View and selecting the LGA Sydney circle to (C) the \textit{Indented Tree-Matrix Comparison View} for further observation. (D)-(F) and (E)-(G) provide two different subcases of predicted testing capacities with updated clinic features.} 
\label{fig:Lenscase2}
\end{figure*}

The LGA of Sydney, which includes the central business district and close surrounds, is a significant place in the Greater Sydney Area. It is an artery of communication to northern Sydney, the southern airport, the eastern coastal beaches, and the western residential areas. Zooming into the Greater Sydney Area, as shown in Fig. \ref{fig:Lenscase2}(A), we drew a polygon to highlight the Sydney LGA and saved the current navigation results to the Storage View. Seven clinics are located in this LGA (see Fig. \ref{fig:Lenscase2}(B)), ranking fourth in the total number of tests performed during this time period. Clicking on the LGA Sydney circle, we then browsed all the clinics’ features in the \textit{Indented Tree-Matrix Comparison View}. Here, we discovered that all clinics allowed walk-ins and all were wheelchair accessible. Only one clinic required a booking before tests. Five out of the seven clinics were open seven days a week and for more than eight business hours per day. The other two only operated on weekdays and only for 4.5 hours per day and 8 hours per day, respectively, as indicated in Fig. \ref{fig:Lenscase2}(C). Looking to check for the impact of these clinics on Sydney’s testing capacities, we delved further into these two subcases. 

\textbf{Exploring the impact of a clinic’s business hours.} Initially, we attempted to slightly adjust the business hours of each clinic, helped by the RF model. The predictions that followed showed that this feature can indeed affect a region’s test capacities. Test numbers are positively correlated with business hours and inversely correlated with break hours. Additionally, over several iterative adjustments to the business hours, we observed that test capacities were not sensitive to sudden increases or decreases in the operating hours on a certain day. Rather, because there is uncertainty in the released figures, the regression model adds together the testing numbers of the previous few days. In addition, we also found that simply changing the business hours of a certain clinic affected the predicted test numbers of other clinics on the same day. In fact, any changes to the business hours on a certain day would cause variations in test counts over a period of 7 days. As an example, to represent this phenomenon, we changed our two weekday-only clinics, Central and 4Cyte, to open on the weekend with the same hours (see Fig. \ref{fig:Lenscase2}(D)), while keeping the clinics’ other objective factors unchanged. Fig. \ref{fig:Lenscase2}(F) shows the resulting predictions. 

Within Fig. \ref{fig:Lenscase2}(F), Panel F1 shows the daily test numbers for both the ground truth and the predictions for 3 Dec 2021 to 31 Dec 2021. Panels F2 and F3 show alternate views of these data. Overall, these results confirm the fact that extending these clinics’ business hours to Saturday and Sunday changes the predicted test numbers for each clinic not only on those days but also for the previous few days. Additionally, from the detailed tooltips, we discovered that both these clinics were significantly more affected (positively) in terms of test number predictions than any of the other clinics in the marked zone on the day of 11 Dec 2021. Generally speaking, opening on the weekend seemed to have a positive impact on the clinics from Thursday to Saturday, most evidently on Saturday, and a negative impact from Monday to Wednesday. However, from a periodic perspective, extending the business hours to the weekend had a positive impact on overall test amounts for the week. 

Panels F2 and F3 also reveal some other interesting patterns. For instance, test numbers in the week of 11 Dec 2021 are different from the following week in that extending the business hours had a greater impact on test numbers this week than the next. The predicted peak was postponed by one day and, even though the clinic’s test capacities would have already reached the regional limit, the highest peak of the predicted values versus the actual values are essentially the same. This indicates that the RF model learned from the time series features in the training dataset that the test numbers during the Christmas and New Year holidays were not sensitive to a clinic’s business hours. As such, the prediction results for the following week remained fairly stable compared to the ground truth. 

\textbf{How different objective factors affect test capacities.} Through multiple progressive explorations, we discovered that only a few clinics supported both walk-ins and drive-through testing. The "Walk-in Allowed" clinics were mostly located in densely populated areas, while the "Drive-through Allowed" clinics were mostly located in remote areas, some of which required booking before tests. Additionally, we learned that test numbers were correlated to both of these factors. In other words, because most clinics were either walk-in or drive-through, not both, different combinations of factors would cause different variations in the testing capacities predicted. Thus, we found that every clinic needed to set a "Yes" to at least one of "Walk-in Allowed" or "Drive-through Allowed", plus a "Yes" to "Referral Required", "Age Limit", "Booking Required", or "Wheelchair Accessible". Not doing this had a negative effect on test counts, and vice versa. Further, we investigated the effects of adding break hours to the clinics’ business hours. The results show that adding a few break hours (say, 1-2 hours) has no discernible impact on testing capacities. 

However, what did intricately impact the predictions was changing confluences of the clinic features. As an example, Fig.\ref{fig:Lenscase2} shows a case where we set "Age limit" and "Requires Booking" for both the Central Clinic and 4Cyte to "Yes" (see Fig.\ref{fig:Lenscase2}(E)). Additionally, we adjusted the business hours to include an hour break each weekday and extended the Central clinic’s weekend hours by 2.5 hours and the 4Cyte clinic’s hours by 4 hours each day. The predictions, shown in Fig.\ref{fig:Lenscase2}(G), are located in the same time period as the previous subcases presented in curve-line mode. Although some trends in both the initial and updated predictions are staggered, there were not too many differences in the overall testing counts for the week because most trends matched properly. However, the pattern shown in Fig.\ref{fig:Lenscase2}(G1) illustrates the differences in trend over a five-week period from 3 Dec 2021 to 7 Jan 2022. Inspecting the values more closely using the tooltips, we discovered the predicted test numbers each day for the first three weeks were generally lower than the initial predictions. This situation can perhaps more clearly be seen from a side-by-side comparison of Fig.\ref{fig:Lenscase2}(F3) and (G1), where the updated clinic features cause a negative influence that outweighs the benefits of extending the weekend business hours in terms of testing capacity. Moreover, Fig.\ref{fig:Lenscase2}(G2) shows us that updating the features for our two focus clinics, Central and 4Cyte, affected the forecast testing numbers for all clinics in the Sydney LGA on certain days. However, the most major effects were felt in these two clinics. This result is consistent with our findings from the previous case study. We also found it convincing that the predicted test numbers changed in the first three weeks before the holidays and then remained relatively constant for the next two weeks. This indicates that the models are accurately reflecting people’s testing behaviors. 

\subsection{Case Study III: Exploring the Optimization Strategy with LGA-based Testing Capacities under Uncertainties}
\begin{figure*}[tb]
  \centering
  \includegraphics[width=\linewidth]{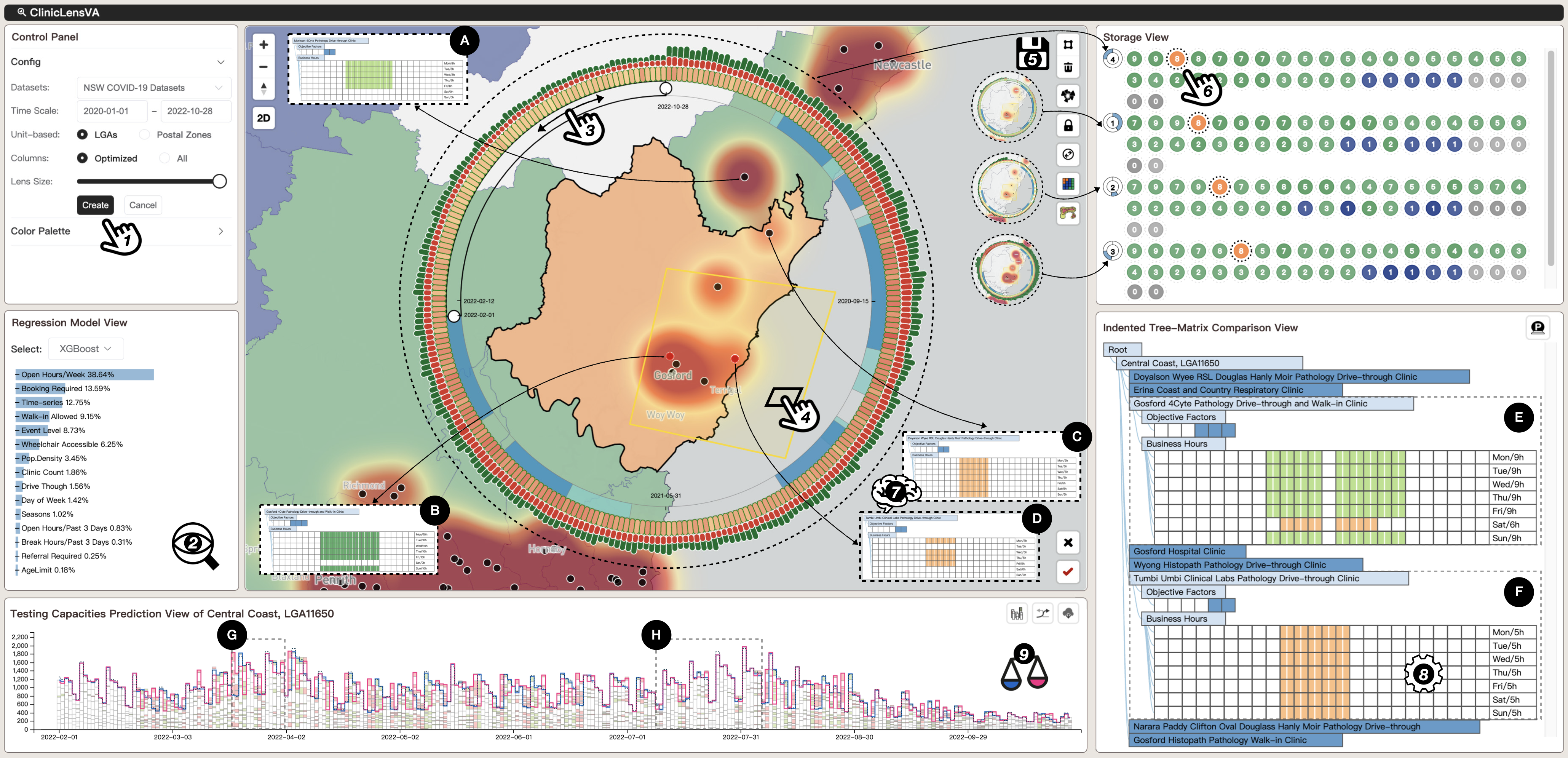}
  \caption{Case Study III: (1) A view of the Control Panel with the time scale set to 1 Jan 2020 to 28 Oct 2022. The unit is set to LGAs; the feature columns to optimized; and the lens size to maximum. (2) Selecting XGBoost and inspecting the importance of each of the features in the Regression Model View after driving the lens to include the three major city scopes in NSW (Sydney, Newcastle, and the Central Coast). (3) Filtering the time periods with sliders in the \textit{Map-Lens View}. (4) Drawing a polygon to highlight the Central Coast LGA. (5) Saving the four time periods as sequences in the Storage View. (6) Focusing on the highlighted Central Coast LGA circle to show the clinic features in the \textit{Indented Tree-Matrix Comparison View}. Comparing the initial features of the clinics (7)  with the updated features (8). (9) The \textit{Testing Capacities Prediction View}, which visualizes the predictions and detailed comparisons of the daily test capacities against the ground truth, the initial predictions, and the updated predictions.}
  \label{fig:teaser}
\end{figure*}

While exploring the aggregated COVID-19 dataset, we realized that making reasonable updates to the clinics’ features could reduce the pressure on certain periods of daily test capacities. Additionally, maintaining a similar volume of tests seemed to achieve the most sustainable balance between clinics' testing volume tolerance and efficient testing services. Thus, we sought to explore patterns and optimize the clinics’ features to prevent the spread of cases and inform policies for coexisting with the virus in the post-COVID era. We discussed the previous two case studies with our domain experts before undertaking the third case study, and, with their feedback in mind, we set out to find strategies for optimizing the testing capacities of the clinics across one LGA by adjusting their clinic factors. This included considering the geographical location of each clinic, their business hours, and their other service details. The complete process is shown in Fig. \ref{fig:teaser}. Our focus points for this case study were three main city scopes in NSW: the LGAs' of Greater Sydney; the Central Coast, a regional area just north of Greater Sydney, and the third most populous region of NSW; and the city of Newcastle, north of the Central Coast, which is NSW’s second most populous region (see Fig. \ref{fig:teaser}(1) and (2)). Fig. \ref{fig:teaser}(3) and (4) show the polygon drawn to highlight the Central Coast LGA. We explored and stored four time periods, as shown in Fig. \ref{fig:teaser}(5) before finally opting for the period from 1 Feb 2022 to 28 Oct 2022. In this period, there were no government interventions, while positive cases remained consistently high. In fact, the positive rates did not start to decline until the end of August which accordingly reflected on the daily test that counts reflect a similar trend with the average weekly total of tests performed reaching \textit{40k} prior to September and eventually dropping to approximately \textit{20k} after.  

As Fig. \ref{fig:teaser}(6) shows, the Central Coast LGA and its eight clinics reported the third highest test counts in the state during this period. We first examined the average testing capacities of each clinic in the Central Coast LGA through the heatmap. Next, we selected this LGA and gathered statistics on the features of the clinics from the \textit{Indented Tree-Matrix Comparison View}. Fig. \ref{fig:teaser}(7) shows the clinics’ features.  

From this Fig.\ref{fig:teaser}(7), we can see that the Central Coast boasts one "almighty" clinic, namely Gosford 4Cyte Pathology Clinic (B), which is open 10 hours per day and supports both walk-in and drive-through clients. It is wheelchair accessible and does not require referrals or bookings. Further, there is no age limit for the patients tested. The remaining seven clinics in this LGA either support walk-ins or drive-throughs, not both, and most operate for only 8 hours per day. From their locations on the \textit{Map-Lens View}, we observed that the Gosford clinic was grouped with three other adjacent clinics that could conduct tests for each other. The other four were located in relatively independent regions.  

Our formulated strategy for optimizing the testing capacities in this LGA is shown in Fig.\ref{fig:teaser}(8). We made the decision to keep the entire business hours the same by appearing the Saturday hours to six hours and appending one hour of break time on the other days, as found in (E). From the \textit{Map-Lens View}, we observed that the Doyalson clinic (C) was geographically close to the Morisset clinic (A) in another LGA (Lake Macquarie) and that Morriset had stronger average test capacities but did not operate on weekends. Hence,  the Doyalson Clinic, to some extent, appeared to be an alternative to Morriset for weekend testing. Both clinics were checked as "Drive-through Allowed" and "Wheelchair Accessible" but, we ruled out setting Doyalson Clinic as Walk-in Allowed by checking its location and surroundings on Google Maps with an abundance of caution. Thus, we did not change any of Doyalson’s features. We further discovered that the Trubi Umbi Clinical Lab (D), which was located in a populated area, was open for 5 hours for all weeks but only provided 5 hours on the COVID-19 test on Monday, Wednesday, Thursday, and Friday in our database. After confirming on HealthDirect \footnote{A government-funded service updated clinic information in real-time: https://www.healthdirect.gov.au/} that the hours for Trubi Lab were not an error in our database, we included updating (D) to (F) as suggestions in the optimization strategy. 

Fig. \ref{fig:teaser}(9) shows the updated predictions. However, although there appeared to be differences between the initial and the updated predictions, it was clear that roughly half of the predicted daily test numbers were nearly identical for the time period, as shown more clearly in (G). At the same time, we discovered that the initial and updated predictions may be staggered during a period of trend. The total forecast test counts in the updated predictions were slightly higher than the counts for the initial predictions in certain weeks, as shown in (H). In other words, the updated predictions were higher than the initial predictions for most days from 19 Mar 2022 to 1 Aug 2022.  

Exploring deeper, we hovered over each clinic bar to read the tooltips and identify which clinics contributed the most significant test amounts to the updated predictions on specific days. The statistics reveal that the Gosford and Tumbi clinics were primarily responsible for the increases. This positive result accords with our strategy of extending Tumbi’s business hours to include Tuesdays, Saturdays, and Sundays and increasing Gosford’s business hours from no hours to 6 hours on Saturdays. Notably, the one-hour break set at the Gosford clinic each day did not cause any unexpected fluctuations. Overall, this proves that \textit{ClinicLens} can be used to develop a reliable strategy for optimizing testing capacities across an LGA. 

\subsection{Expert Interview}
After conducting the case studies, we arranged one-on-one structured interviews with our three domain experts (E1-E3). From a discussion of our findings from the three case studies, we received affirmation from the domain experts that \textit{ClinicLens} archived their initial expectations. We then provided them with tutorials and encouraged them to use \textit{ClinicLens} independently to explore issues of their own interest. Their feedback and qualitative preferences are summarized below. 

\textbf{Reliability of the Back-end Engine.} According to all the experts, \textit{ClinicLens} provides productive results when exploring and optimizing clinic testing capacities given uncertainty. All the information included in the Back-end Engine was reasonable and, after selecting the appropriate regression model based on performance, the experts felt confident in performing their analysis tasks. This was, in part, thanks to \textit{ClinicLens}’s assistance in providing the RF and XGBoost models with automatic feature extraction as steady streams to the Front-end Visualization. Additionally, \textit{ClinicLens} also offers all-feature regression, which means that new feature inputs can easily be added in the future.

\textbf{Effectiveness of the Front-end Visualization.} The consensus among all domain experts was that each view in the Front-end Visualization was considered in terms of its visual form. They were all impressed with the \textit{ClinicLens} design. E1 commented on \textit{ClinicLens} overall as being a valuable visual analytics system. He was especially impressed with the visual expressions in the \textit{Testing Capacities Prediction View}. E2 highlighted that \textit{"including a mobile lens on the map is a smart move. In addition to representing rich information, such as combining interventions, the \textit{Map-Lens View} can be persuasively explored using several different interaction techniques and serves as a trigger for other views for collaborative exploration".} E3, on the other hand, particularly admired the \textit{Indented Tree-Matrix Comparison View}. He thought the indentations of the statistics not only effectively conveyed the structure of the data in a constrained area but also allowed the user to adjust the clinic features for prediction in a worry-free manner. Cycling through the user-friendly interactions in each view, the experts agreed that \textit{ClinicLens} had a low learning curve and could offer quick responses to COVID-19-related questions through visual analytics and machine learning techniques that were not just limited to clinic testing capacities. In general, they felt the whole visual pipeline ran smoothly over the COVID-19 aggregated dataset, and the experts thought it could significantly increase the efficiency of their analyses.

\section{Discussion, Limitation, and Future Work}
% This section assembles the experience learned from exploring and optimizing test capacities.

\textbf{The purpose of exploring testing capacities.} The first and most important thing to discuss is that we must accept the fact that predictions inevitably contain bias or variance \cite{dst_Cao15}. Any slight change in a  clinic’s factors might lead to changes in testing capacities. The predictions may vary with different attempts to change the same feature, but few, if any, changes will result in a change of large magnitude. Further, some changes resulted in staggered variations between the initial and updated prediction results. This and the inherent uncertainty in the data are the primary reasons we recapped. We cannot trace the total test numbers announced for a certain day to the actual tests conducted by each clinic. Rather, we can only estimate a clinic’s test capacity based on its given features with machine learning models. However, this does offer a good overview from a visual analytics perspective. Further, our aim stretches beyond simply predicting test capacities. We have designed \textit{ClinicLens} to help users optimize the testing capacity of a clinic, a postcode, an LGA, or an entire state by giving them the option to adjust multiple different clinic features.  

\textbf{Optimization Strategies.} There is no limit to the number of optimization strategies that can be explored through \textit{ClinicLens}. Many strategies formulated will increase the testing capacity of the selected region. However, considering that there is always bias and uncertainty in predictions, we must experiment iteratively and combine real situations with actual experience to find the most appropriate strategies. Generally speaking, a few solid and preferred strategies should ultimately be reached after observing and closely examining a given situation. Our case studies inspired and motivated the domain experts to devise their own feasible strategies. For example, Case Study II taught them that they could tweak various combinations of clinic factors, including just one clinic’s business hours, to create a positive effect on a whole region’s testing capacity. Case Study III showed them that ensuring each clinic was open consistently during the same business hours each day also increased testing capacities, whereas adding break hours to days had little impact. 

\textbf{Generalizability and Scalability.} The generalizability and scalability of our \textit{ClinicLens} can be considered in terms of three perspectives. First, the aggregated COVID-19 dataset is only an example of the data that could be used to power \textit{ClinicLens}. Our demonstrated visualizations ran perfectly both over subsegments of the data and over the whole time period under study. Second, the architecture of \textit{ClinicLens} is designed to purposefully separate the Back-end Engine from the Front-end Visualization. This means that different models can be added or revised very easily. In addition, the framework provides all-feature regression, which means that \textit{ClinicLens} is flexible. It can easily accommodate future additions of new features or input. Lastly, even though \textit{ClinicLens} in its demonstration form is limited to exploring COVID-19 data, all our domain experts agree that this could be a productive visual analytics system for other epidemiological analyses. With slightly adjusted parameters and by adding a few more visualization views to help analyze spatiotemporal features, \textit{ClinicLens} could be used to analyze clinic testing capacities for other infectious diseases. 

\textbf{Limitations.} Our approach remains constrained by certain implicit features of the clinic information. First, some implicit features may cause fluctuations in a clinic’s predicted test capacity. For example, uneven population density and the size of each clinic in terms of the number of employees and the quality of the medical services provided may influence a person’s preference for where they get tested. Additionally, we did not account for any differences in business hours that may have impacted the prediction results. For example, there would clearly be different forecasts for clinics that were open at 5 am and not 5 pm. We also did not demarcate clinics authorized to conduct rapid testing for international airline departures (where the results are made available within 48 hours) as these features give rise to more complex regression analysis. One feature that the experts requested after using \textit{ClinicLens} was the ability to add or delete clinics from an area and forecast the impact on the other clinics in the region. \textit{ClinicLens} does currently not do this, but it may be a function we add in future work. 

\textbf{Future work.} We will consider attaching more implicit features and applying state-of-the-art algorithms to the Back-end Engine, such as the algorithm outlined in \cite{liu2022modeling} to make the back-engine even more robust to uncertainty and to enhance its feature modeling and predictions. We may also develop a clinic editor as a new function. Additionally, we have plans to expand \textit{ClinicLens} to other states and territories in Australia to create a real-time system for managing future outbreaks of infectious diseases.

\section{Conclusion}
In this paper, we presented \textit{ClinicLens}, an interactive visual analytics system for use by domain experts to both explore location-based COVID-19 case and test statistics as well as optimize the testing capacities of the clinics in selected locations. Motivated by the current challenges in preparing for infectious disease outbreaks and informed by expert requirements, \textit{ClinicLens} comprises Back-end Engine that is automatically driven by AI to identify and extract features that may affect clinic testing capacities and Front-end Visualization system that offers multiple perspectives on the data to support planning and decision-making. Importantly, the framework is robust to the uncertainty inherent in the available COVID-19 datasets. Three real-world case studies along with expert interviews validate the usefulness and effectiveness of \textit{ClinicLens}. As such, we believe \textit{ClinicLens} offers a fresh perspective on the decision-making surrounding clinic testing capacities for infectious disease management. In the future, we intend to address the current limitations of \textit{ClinicLens} as well as deliver more visual assistance for real-world analysis tasks to domain experts.

%\bibliographystyle{abbrv}
% \bibliographystyle{abbrv-doi}
%\bibliographystyle{abbrv-doi-narrow}
%\bibliographystyle{abbrv-doi-hyperref}
%\bibliographystyle{abbrv-doi-hyperref-narrow}

% or


% \appendices

% % you can choose not to have a title for an appendix
% % if you want by leaving the argument blank
% \section{}

\ifCLASSOPTIONcompsoc
  \section*{Acknowledgments}
\else
  \section*{Acknowledgment}
\fi

The authors would like to thank all of the domain experts from the Australian government who contributed their expertise, insightful comments, and feedback, and appreciate the work of the proofreaders.

\ifCLASSOPTIONcaptionsoff
  \newpage
\fi



\bibliographystyle{IEEEtran}
\bibliography{main}
% \bibliography{IEEEabrv,../bib/paper}

% H.~Kopka and P.~W. Daly, \emph{A Guide to {\LaTeX}}, 3rd~ed.\hskip 1em plus

% \end{thebibliography}

% or if you just want to reserve a space for a photo:、

\newpage
\begin{IEEEbiography}[{\includegraphics[width=1in,height=1.25in,clip,keepaspectratio]{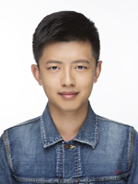}}]{Yu Dong} received his M.S. degree from the Beijing Technology and Business University in 2017. He is currently a Ph.D. candidate at the University of Technology, Sydney. His research interests include data visualization and visual analytics, machine learning, and human-computer interaction. 
\end{IEEEbiography}

\begin{IEEEbiography}[{\includegraphics[width=1in,height=1.25in,clip,keepaspectratio]{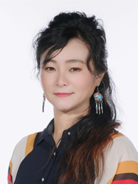}}]{Jie Liang} received her doctorate in data visualization from the University of Technology, Sydney, and now leads the Data Visualization Research Lab in the Visualization Institute at UTS. Her research interests focus on data visualization and visual analytics.
\end{IEEEbiography}

\begin{IEEEbiography}[{\includegraphics[width=1in,height=1.25in,clip,keepaspectratio]{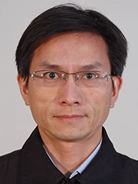}}]{Longbing Cao} has earned a Ph.D. degree in pattern recognition and intelligent systems and another in computing sciences. He is a professor and ARC Future Fellow (Level 3) with the University of Technology Sydney. His research interests include AI, data science, analytics and machine learning, behavior informatics, and enterprise applications. He has more than 300 publications and is a senior member of IEEE.
\end{IEEEbiography}

\begin{IEEEbiography}[{\includegraphics[width=1in,height=1.25in,clip,keepaspectratio]{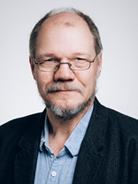}}]{Daniel Catchpoole} is a professor at the University of Technology Sydney and head of the Biospecimens Research Services within Kids Research at The Children’s Hospital at Westmead. His research interests lie in the application of high-end computational analysis and visual analytics to complex multi-dimensional data to draw out actionable knowledge that has clinical relevance and can be used to guide public health.
\end{IEEEbiography}

%\newpage



%\vfill


\end{document}